\documentclass[english]{article}
\usepackage[T1]{fontenc}
\usepackage[latin9]{inputenc}
\usepackage{float}
\usepackage{amsmath}
\usepackage{amssymb}
\usepackage{graphicx}

\makeatletter
\newcommand{\lyxaddress}[1]{
	\par {\raggedright #1
	\vspace{1.4em}
	\noindent\par}
}

\makeatother

\usepackage{babel}
\begin{document}
\title{\textbf{ON THE GENERALIZED LEMAITRE TOLMAN BONDI METRIC: CLASSICAL
SENSITIVITIES AND QUANTUM EINSTEIN-VAZ SHELLS}}
\author{\textbf{$^{1}$Mohammadreza Molaei and $^{2}$Christian Corda}}
\maketitle

\lyxaddress{$^{1}$Department of Mathematics, Shahid Bahonar University of Kerman,
Kerman,, Iran. E-mail: $mrmolaei@uk.ac.ir$}

\lyxaddress{$^{2}$SUNY Polytechnic Institute, 13502 Utica, New York, USA, Istituto
Livi, 59100 Prato, Tuscany, Italy and International Institute for
Applicable Mathematics and Information Sciences, B. M. Birla Science
Centre, Adarshnagar, Hyderabad 500063, India. E-mail: $cordac.galilei@gmail.com$}
\begin{abstract}
In this paper, in the classical framework we evaluate the lower bounds
for the sensitivities of the generalized Lemaitre Tolman Bondi metric.
The calculated lower bounds via the linear dynamical systems $L_{\frac{\partial}{\partial\theta}}$,
$L_{\frac{\partial}{\partial r}}$, and $L_{\frac{\partial}{\partial\phi}}$
are $-\ln2+\ln|(\dot{R}B)^{2}-(R')^{2}|-2\ln|B|,$ $2\ln|\dot{B}|-\ln2$
and $-\ln2-2\ln|B|+\ln|(\dot{R}^{2}B^{2}-R'^{2})\sin^{2}\theta-B^{2}\cos^{2}\theta|$
respectively. We also show that the sensitivities and the lower sensitivities
via $L_{\frac{\partial}{\partial t}}$ are zero. In the quantum framework
we analyse the properties of the Einstein-Vaz shells which are the
final result of the quantum gravitational collapse arising from the
Lemaitre Tolman Bondi discussed by Vaz in 2014. In fact, Vaz showed
that continued collapse to a singularity can only be obtained if one
combines two independent and entire solutions of the Wheeler-DeWitt
equation. Forbidding such a combinatin leads naturally to matter condensing
on the Schwarzschild surface during quantum collapse. In that way,
an entirely new framework for black holes (BHs) has emerged. The approach
of Vaz as also consistent with Einstein's idea in 1939 of the localization
of the collapsing particles within a thin spherical shell. Here, following
an approach of oned of us (CC), we derive the BH mass and energy spectra
via a Schrodinger-like approach, by further supporting Vaz's conclusions
that instead of a spacetime singularity covered by an event horizon,
the final result of the gravitational collapse is an essentially quantum
object, an extremely compact ``dark star''. This ``gravitational atom''
is held up not by any degeneracy pressure but by quantum gravity in
the same way that ordinary atoms are sustained by quantum mechanics.
Finally, we discuss the time evolution of the Einstein-Vaz shells. 
\end{abstract}
\begin{quote}
\end{quote}

\section{INTRODUCTION}

The metric 
\[
ds^{2}=dt^{2}-D^{2}(t)dr^{2}-E^{2}(t)[d\theta^{2}+F_{l}^{2}(\theta)d\psi^{2}]
\]
 has been introduced first by Shamir \cite{key-11} to describe the
relation between Kantowski-Sachs and Bianchi type cosmological models.
In this metric $l$ is the spatial curvature index, and $D$, $E$,
and $F_{l}$ are functions of $t$, and $\theta$ respectively. In
fact, if we take $l=1$ and $F_{1}(\theta)=\sin(\theta)$, then we
have Kantowski-Sachs model \cite{key-6}, which is a space-time with
an anisotropic background. In the case $l=0$ and $F_{0}(\theta)=\theta,$
we have locally rotationally symmetric Bianchi type-I model \cite{key-1},
and in the case $l=-1$ and $F_{-1}(\theta)=\sinh(\theta),$ we have
locally rotationally symmetric Bianchi type-III cosmological model.
A generalization of Shamir's metric is the general form of the Lemaitre
Tolman Bondi metric (or LTB metric briefly) \cite{key-3,key-7,key-13}.
This general form in the coordinate $(t,r,\theta,\phi)$ is 
\[
ds^{2}=-dt^{2}+B^{2}(t,r)dr^{2}+R^{2}(t,r)d\theta^{2}+R^{2}(t,r)\sin^{2}\theta d\phi^{2},
\]
where $B(t,r)$ and $R(t,r)$ are two positive functions \cite{key-5}.
This metric appears in the consideration of gravitational collapse,
the cosmic censorship \cite{key-4,key-8}, and quantum gravity \cite{key-2,key-12,key-14,key-16}.
The notion of sensitivity for a non Riemannian metric has been introduced
first in 2016 \cite{key-9}. In fact, the sensitivity of a metric
determines an upper bound for the deviation of it from the Riemannian
case. The lower bound of it's deviation from the Riemannian case is
called the lower sensitivity of it \cite{key-10}. In this paper we
evaluate the lower sensitivity and the sensitivity of the general
form of LTB metric. We prove that in the direction of $\frac{\partial}{\partial t}$
the sensitivity and the lower sensitivity of LTB metric are zero.
If we choose the direction $\frac{\partial}{\partial\theta},$ then
we show that the sensitivity of LTB metric is grater or equal than
$-\ln2+\ln|(\dot{R}B)^{2}-(R')^{2}|-2\ln|B|.$ For the direction $\frac{\partial}{\partial r},$
we find the lower bound $2\ln|\dot{B}|-\ln2$ for the sensitivity
of LTB metric. We show that the lower bound for the sensitivity in
the direction $\frac{\partial}{\partial\phi}$ is 
\[
-\ln2-2\ln|B|+\ln|(\dot{R}^{2}B^{2}-R'^{2})\sin^{2}\theta-B^{2}\cos^{2}\theta|.
\]
After the above summarized classical analysis, in the quantum framework
we consider Vaz's approach \cite{key-17}, where continued collapse
to a singularity can only be obtained if one combines two independent
and entire solutions of the Wheeler-DeWitt equation. By forbidding
such a combination \cite{key-17}, one gets a natural result of matter
condensing on the apparent horizon during quantum collapse. In that
way, an entirely new BH framework has emerged \cite{key-17}. The
approach of Vaz was also consistent with Einstein's idea in 1939 of
the localization of the collapsing particles within a thin spherical
shell \cite{key-18}. Following \cite{key-19}, the BH mass and energy
spectra via a Schrodinger-like approach will be obtained. This further
supports Vaz's conclusions that instead of a spacetime singularity
covered by an event horizon, the final result of the gravitational
collapse is an essentially quantum object, an extremely compact ``dark
star''. This ``gravitational atom'' is held up not by any degeneracy
pressure, but by quantum gravity in the same way that ordinary atoms
are sustained by quantum mechanics. Finally, we discuss the time evolution
of the Einstein-Vaz shells \cite{key-20}. 

\section{SENSITIVITIES IN THE DIRECTIONS $\frac{\partial}{\partial t}$ and
$\frac{\partial}{\partial r}$ }

We begin this section by a short overview on the notion of sensitivity
of a metric based on a chart (local coordinate). We assume $\nabla$
is the Levi-Civita connection corresponding to a metric $g$ on a
manifold $M$ \cite{key-15}. Thus, the components of $\nabla$ in
a chart $(U,x^{i})$ are: 
\begin{equation}
\Gamma_{ij}^{k}=\frac{1}{2}g^{km}(\frac{\partial}{\partial x^{i}}(g_{jm})+\frac{\partial}{\partial x^{j}}(g_{im})-\frac{\partial}{\partial x^{m}}(g_{ij})).\label{eq2}
\end{equation}
If $p\in U$, and $i\in\{1,2,\cdots,m\}$, then the linear map $L_{\frac{\partial}{\partial x^{i}}}:T_{p}M\rightarrow T_{p}M$
is defined by 
\[
L_{\frac{\partial}{\partial x^{i}}}(\frac{\partial}{\partial x^{j}}(p))=\Gamma_{ij}^{k}(p)\frac{\partial}{\partial x^{k}}(p).
\]
For a natural number $n$, and $i\in\{1,2,\cdots,m\},$ the functions
$\gamma_{n,\frac{\partial}{\partial x^{i}}}$ on ${\bf R}$ and $\lambda_{n,\frac{\partial}{\partial x^{i}}}$
on $T_{p}M$ are defined by 
\[
{\gamma_{n,\frac{\partial}{\partial x^{i}}}}=\inf_{\vert v\vert\leq\dfrac{1}{2^{n}}}\lambda_{n,\frac{\partial}{\partial x^{i}}}(v),
\]
where $\lambda_{n,\frac{\partial}{\partial x^{i}}}(v)=g(L_{\frac{\partial}{\partial x^{i}}}^{n}(v),L_{\frac{\partial}{\partial x^{i}}}^{n}(v))$.
If we define $R_{n,\frac{\partial}{\partial x^{i}}}$ by 
\[
R_{n,\frac{\partial}{\partial x^{i}}}=\left\{ \begin{array}{rl}
 & \dfrac{ln\vert\gamma_{n,\frac{\partial}{\partial x^{i}}}\vert}{n}~~~if~~~\gamma_{n,\frac{\partial}{\partial x^{i}}}\notin\{0,-\infty\}\\
 & 0~~~~~~~~~~~~~~~~~~~~~~\hspace{0.5cm}otherwise
\end{array},\right.
\]
then, the lower sensitivity of $g$ on $U$ in the direction of $\frac{\partial}{\partial x^{i}}$
is the function 
\[
\mathbb{H}_{g}(\frac{\partial}{\partial x^{i}}):U\rightarrow[-\infty,\infty]
\]
defined by 
\[
\mathbb{H}_{g}(\frac{\partial}{\partial x^{i}})(p)={\displaystyle \liminf_{n\rightarrow\infty}R_{n,\frac{\partial}{\partial x^{i}}}.}
\]
If in the former equality we replace $\liminf$ with $\limsup$, then
the resulted function is denoted by $\mathbb{H}^{g}(\frac{\partial}{\partial x^{i}})$,
and it is called the sensitivity of $g$ on $U$ in the direction
of $\frac{\partial}{\partial x^{i}}$. In the basis $\{\frac{\partial}{\partial t},\frac{\partial}{\partial r},\frac{\partial}{\partial\theta},\frac{\partial}{\partial\phi}\},$
the computed non-zero Christoffel symbols of the Levi-Civita connection
corresponding to the LTB metric are: 
\begin{equation}
\Gamma_{21}^{2}=\Gamma_{12}^{2}=\frac{\dot{B}}{B},~\Gamma_{22}^{2}=\frac{B'}{B},~\Gamma_{22}^{1}=B\dot{B},~\Gamma_{41}^{4}=\Gamma_{31}^{3}=\Gamma_{14}^{4}=\Gamma_{13}^{3}=\frac{\dot{R}}{R}\label{eq0}
\end{equation}
\[
\Gamma_{42}^{4}=\Gamma_{32}^{3}=\Gamma_{24}^{4}=\Gamma_{23}^{3}=\frac{R'}{R},~\Gamma_{33}^{1}=R\dot{R},~\Gamma_{33}^{2}=-\frac{RR'}{B^{2}},
\]
\[
\Gamma_{44}^{3}=-\sin(\theta)\cos(\theta),~\Gamma_{43}^{4}=\Gamma_{34}^{4}=\cot\theta,
\]
 
\[
\Gamma_{44}^{2}=-\frac{RR'\sin^{2}(\theta)}{B^{2}},~\Gamma_{44}^{1}=R\dot{R}\sin^{2}(\theta),
\]
where $\dot{B}=\frac{\partial B}{\partial t}$, $\dot{R}=\frac{\partial R}{\partial t}$,
$B'=\frac{\partial B}{\partial r}$, and $R'=\frac{\partial R}{\partial r}$.
The sensitivity and the lower sensitivity of LTB metric in the direction
of $\frac{\partial}{\partial t}$ are zero, and in the points $(t,r,\theta,\phi)$
with $\dot{B}\neq0$, we have $\mathbb{H}^{g}(\frac{\partial}{\partial r})(t,r,\theta,\phi)\geq2\ln|\dot{B}|-\ln2$.
The matrix of $L_{\frac{\partial}{\partial t}}$ in the basis $\{\frac{\partial}{\partial t},\frac{\partial}{\partial r},\frac{\partial}{\partial\theta},\frac{\partial}{\partial\phi}\}$
is 
\[
A=\left(\begin{array}{cccc}
0 & 0 & 0 & 0\\
0 & \frac{\dot{B}}{B} & 0 & 0\\
0 & 0 & \frac{\dot{R}}{R} & 0\\
0 & 0 & 0 & \frac{\dot{R}}{R}
\end{array}\right).
\]
Thus, 
\[
\lambda_{n,\frac{\partial}{\partial t}}(v^{1},v^{2},v^{3},v^{4})=\frac{\dot{B}^{2n}}{B^{2n-2}}(v^{2})^{2}+\frac{\dot{R}^{2n}}{R^{2n-2}}(v^{3})^{2}+\frac{\dot{R}^{2n}}{R^{2n-2}}\sin^{2}(\theta)(v^{4})^{2}.
\]
Therefore, $\gamma_{n,\frac{\partial}{\partial t}}$ is the constant
zero function, and so $R_{n,\frac{\partial}{\partial t}}=0$. Thus,
$\mathbb{H}_{g}(\frac{\partial}{\partial t})=\mathbb{H}^{g}(\frac{\partial}{\partial t})=0$.
The matrix of $L_{\frac{\partial}{\partial r}}$ in the basis $\{\frac{\partial}{\partial t},\frac{\partial}{\partial r},\frac{\partial}{\partial\theta},\frac{\partial}{\partial\phi}\}$
is 
\[
D=\left(\begin{array}{cccc}
0 & B\dot{B} & 0 & 0\\
\frac{\dot{B}}{B} & \frac{B'}{B} & 0 & 0\\
0 & 0 & \frac{R'}{R} & 0\\
0 & 0 & 0 & \frac{R'}{R}
\end{array}\right).
\]
The entry $(1,1)$ of $D^{n}$ determines the sensitivity of $L_{\frac{\partial}{\partial r}}$.
By using of Maple we see that the entry $(1,1)$ of the matrix $D^{2n}$
for $n>1$ has the form 
\[
\dot{B}^{2n}+a_{1}(\frac{B'}{B})^{2}\dot{B}^{2n-2}+a_{2}(\frac{B'}{B})^{4}\dot{B}^{2n-4}+\cdots+a_{n-1}(\frac{B'}{B})^{2n-2}\dot{B}^{2},
\]
where $a_{1},\cdots,a_{n-1}$ are fixed natural numbers. For example,
for $n=5$ we have $a_{1}=10$, $a_{2}=15$, $a_{3}=7$, and $a_{4}=1.$
In this case 
\[
\gamma_{2n,\frac{\partial}{\partial r}}=-2^{-2n}(\dot{B}^{2n}+a_{1}(\frac{B'}{B})^{2}\dot{B}^{2n-2}+a_{2}(\frac{B'}{B})^{4}\dot{B}^{2n-4}+\cdots+a_{n-1}(\frac{B'}{B})^{2n-2}\dot{B}^{2})^{2}.
\]
Thus, $\gamma_{2n,\frac{\partial}{\partial r}}\leq-2^{-2n}\dot{B}^{4n}<0.$
Hence, $R_{2n,\frac{\partial}{\partial r}}\geq\frac{\ln|-2^{-2n}\dot{B}^{4n}|}{2n}=2\ln|\dot{B}|-\ln2$.
Therefore, $\mathbb{H}^{g}(\frac{\partial}{\partial r})(t,r,\theta,\phi)\geq2\ln|\dot{B}|-\ln2$
for all $(t,r,\theta,\phi),$ with $\dot{B}(t,r)\neq0$. We have sketched
the lower bound of $\mathbb{H}^{g}(\frac{\partial}{\partial r})(t,r,\theta,\phi)$
for the case $B=e^{tr}$ in figure 1. 
\begin{figure}[H]

\includegraphics[scale=0.5]{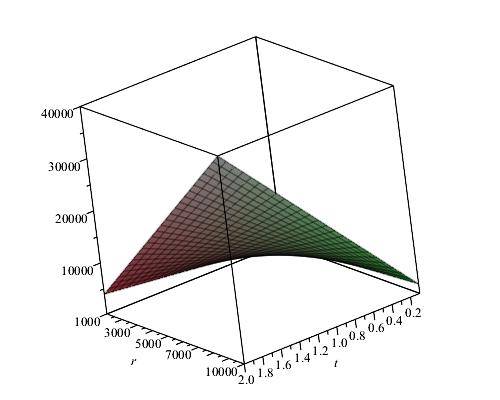}\caption{The graph of the lower bound of $\mathbb{H}^{g}(\frac{\partial}{\partial r})(t,r,\theta,\phi)$,
when $B=e^{tr}$. }

\end{figure}

\section{SENSITIVITIES IN THE DIRECTIONS $\frac{\partial}{\partial\theta}$
and $\frac{\partial}{\partial\phi}$ }

We begin this section by the following theorem. The sensitivity of
LTB metric in the direction of $\frac{\partial}{\partial\theta}$
in the points with $\dot{R}\neq0$ and $R'^{2}\neq(\dot{R}B)^{2}$,
is greater or equal than $-\ln2+\ln|(\dot{R}B)^{2}-(R')^{2}|-2\ln|B|.$
Since $L_{\frac{\partial}{\partial\theta}}(\frac{\partial}{\partial t})=\frac{\dot{R}}{R}\frac{\partial}{\partial\theta}$,
$L_{\frac{\partial}{\partial\theta}}(\frac{\partial}{\partial r})=\frac{{R'}}{R}\frac{\partial}{\partial\theta}$,
$L_{\frac{\partial}{\partial\theta}}(\frac{\partial}{\partial\theta})=R\dot{R}\frac{\partial}{\partial t}-\frac{{RR'}}{B^{2}}\frac{\partial}{\partial r}$,
and $L_{\frac{\partial}{\partial\theta}}(\frac{\partial}{\partial\phi})=\cot\theta\frac{\partial}{\partial\phi}$,
then the matrix of $L_{\frac{\partial}{\partial t}}$ in the basis
$\{\frac{\partial}{\partial t},\frac{\partial}{\partial r},\frac{\partial}{\partial\theta},\frac{\partial}{\partial\phi}\}$
is 
\[
E=\left(\begin{array}{cccc}
0 & 0 & {\dot{R}}{R} & 0\\
0 & 0 & -\frac{RR'}{B^{2}} & 0\\
\frac{\dot{R}}{R} & \frac{R'}{R} & 0 & 0\\
0 & 0 & 0 & \cot\theta
\end{array}\right).
\]
For a natural number $n$, the $(1,1)$ entry of the matrix $E^{2n}$
is $\frac{(\dot{R})^{2}((\dot{R}B)^{2}-(R')^{2})^{n-1}}{B^{2n-2}}.$
Thus $\gamma_{2n,\frac{\partial}{\partial\theta}}\leq-\frac{2^{-2n}\dot{R}^{4}((\dot{R}B)^{2}-(R')^{2})^{2n-2}}{B^{4n-4}}.$
In the points with $\dot{R}\neq0$ and $R'^{2}\neq(\dot{R}B)^{2}$
we have $R_{2n,\frac{\partial}{\partial\theta}}\geq\frac{\ln|-\frac{2^{-2n}\dot{R}^{4}((\dot{R}B)^{2}-(R')^{2})^{2n-2}}{B^{4n-4}}|}{2n}=-\ln2+\frac{4\ln|\dot{R}|}{2n}+(2n-2)\frac{\ln|(\dot{R}B)^{2}-(R')^{2}|}{2n}-(4n-4)\frac{\ln|B|}{2n}.$
Thus $\mathbb{H}^{g}(\frac{\partial}{\partial\theta})(t,r,\theta,\phi)\geq-\ln2+\ln|(\dot{R}B)^{2}-(R')^{2}|-2\ln|B|.$ 

The lower bound of $\mathbb{H}^{g}(\frac{\partial}{\partial\theta})(t,r,\theta,\phi)$
for the case $B=e^{tr}$ and $R=t^{\frac{2}{3}}$ is sketched in figure
2.

\begin{figure}

\includegraphics[scale=0.5]{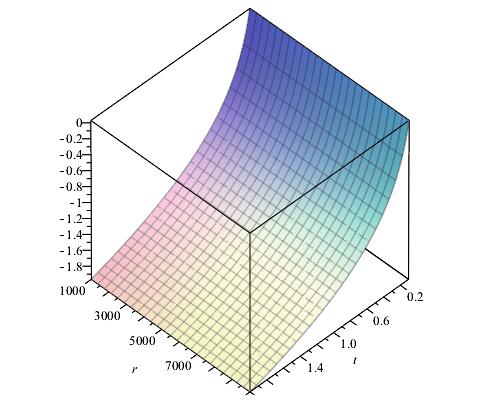}\caption{The graph of the lower bound of $\mathbb{H}^{g}(\frac{\partial}{\partial\theta})(t,r,\theta,\phi)$,
when $B=e^{tr}$ and $R=t^{\frac{2}{3}}$.}

\end{figure}

For the direction $\frac{\partial}{\partial\phi}$ we have the next
theorem. In the points $(t,r,\theta,\phi)$ which $\dot{R}\sin^{2}\theta\neq0$
and $(\dot{R}^{2}B^{2}-R'^{2})\sin^{2}\theta-B^{2}\cos^{2}\theta\neq0,$
we have 
\[
\mathbb{H}^{g}(\frac{\partial}{\partial\phi})(t,r,\theta,\phi)\geq-\ln2-2\ln|B|+\ln|(\dot{R}^{2}B^{2}-R'^{2})\sin^{2}\theta-B^{2}\cos^{2}\theta|.
\]
Since $L_{\frac{\partial}{\partial\phi}}(\frac{\partial}{\partial t})=\frac{\dot{R}}{R}\frac{\partial}{\partial\phi}$,
$L_{\frac{\partial}{\partial\phi}}(\frac{\partial}{\partial r})=\frac{R'}{R}\frac{\partial}{\partial\phi}$,
$L_{\frac{\partial}{\partial\phi}}(\frac{\partial}{\partial\phi})=\cot\theta\frac{\partial}{\partial\phi}$,
and $L_{\frac{\partial}{\partial\phi}}(\frac{\partial}{\partial\phi})=R\dot{R}\sin^{2}\theta\frac{\partial}{\partial t}-\frac{RR'}{B^{2}}\sin^{2}\theta\frac{\partial}{\partial r}-\sin\theta\cos\theta\frac{\partial}{\partial\theta}$.
Then the matrix of $L_{\frac{\partial}{\partial\phi}}$ in the basis
$\{\frac{\partial}{\partial t},\frac{\partial}{\partial r},\frac{\partial}{\partial\theta},\frac{\partial}{\partial\phi}\}$
is 
\[
G=\left(\begin{array}{cccc}
0 & 0 & 0 & {\dot{R}}{R}\sin^{2}\theta\\
0 & 0 & 0 & -\frac{RR'}{B^{2}}\sin^{2}\theta\\
0 & 0 & 0 & -\sin\theta\cos\theta\\
\frac{\dot{R}}{R} & \frac{R'}{R} & \cot\theta & 0
\end{array}\right).
\]
The entry $(1,1)$ of the matrix $G^{2n}$ is: 
\[
\frac{\dot{R}^{2}\sin^{2}\theta}{B^{2n-2}}((\dot{R}^{2}B^{2}-R'^{2})\sin^{2}\theta-B^{2}\cos^{2}\theta)^{n-1}.
\]
So, 
\[
\gamma_{2n,\frac{\partial}{\partial\phi}}\leq-2^{-2n}\frac{\dot{R}^{4}\sin^{4}\theta}{B^{4n-4}}((\dot{R}^{2}B^{2}-R'^{2})\sin^{2}\theta-B^{2}\cos^{2}\theta)^{2n-2}.
\]
Hence, in the points $(t,r,\theta,\phi),$ which $\dot{R}\sin^{2}\theta\neq0$
and $(\dot{R}^{2}B^{2}-R'^{2})\sin^{2}\theta-B^{2}\cos^{2}\theta\neq0,$
we have 
\[
R_{2n,\frac{\partial}{\partial\phi}}\geq\frac{\ln|-2^{-2n}\frac{\dot{R}^{4}\sin^{4}\theta}{B^{4n-4}}((\dot{R}^{2}B^{2}-R'^{2})\sin^{2}\theta-B^{2}\cos^{2}\theta)^{2n-2}|}{2n}=
\]
 
\[
-\ln2+\frac{\ln|\dot{R}^{4}\sin^{4}\theta|}{2n}-\frac{(4n-4)\ln|B|}{2n}+\frac{2n-2}{2n}\ln|(\dot{R}^{2}B^{2}-R'^{2})\sin^{2}\theta-B^{2}\cos^{2}\theta|.
\]
Hence, 
\[
\mathbb{H}^{g}(\frac{\partial}{\partial\phi})(t,r,\theta,\phi)\geq-\ln2-2\ln|B|+\ln|(\dot{R}^{2}B^{2}-R'^{2})\sin^{2}\theta-B^{2}\cos^{2}\theta|.
\]

\section{EINSTEIN-VAZ SCHELLS IN THE QUANTUM FRAMEWORK}

Vaz won the Second Prize in the 2014 Gravity Research Foundation Essay
Competition by realizing a quantum approach to the spherical collapse
of inhomogeneous dust in AdS of dimension $d=n+2$ \cite{key-17},
which is described by the previously analysed LeMaitre Tolman Bondi
family of metrics. The model can be expressed in canonical form after
a series of simplifying canonical transformations and after absorbing
the surface terms \cite{key-17}. By using Dirac quantization of the
constraints leading to a Wheeler-DeWitt equation, two independent
solutions in terms of shell wave functions supported everywhere in
spacetime come out \cite{key-17} (in this Section and in next one,
Planck units will be used, i.e. $G=c=k_{B}=\hbar=\frac{1}{4\pi\epsilon_{0}}=1$)
\begin{equation}
\psi_{i}=\psi_{i}^{\left(1\right)}+A_{i}\psi_{i}^{\left(2\right)}.\label{eq: wave functions}
\end{equation}
Here $\psi_{i}^{\left(1\right)}$ represents dust shells condensing
to the Schwarzschild surface (which becomes an apparent horizon) on
both sides of it and $\psi_{i}^{\left(2\right)}$ represents dust
shells move away from the Schwarzschild surface on either side of
it where the exterior, outgoing wave is suppressed by the Boltzmann
factor at the Hawking temperature for the shell, see \cite{key-17}
for details. It is important to emphasize that there is nothing within
the theory that suggests a value for $A_{i}$ \cite{key-17}. Indeed,
one should need further input in order to determine these amplitudes
\cite{key-17}. If $0<\left|A_{i}\right|\leq1,$ then the dust will
ultimately pass through the horizon via a continued collapse arriving
at a central singularity \cite{key-17}. Consequently, an event horizon
will form, with emission of Hawking radiation in the exterior \cite{key-17}.
In order to avoid the formation of the central singularity, $\left|A_{i}\right|$
must vanish \cite{key-17}. Then, $\psi_{i}^{\left(1\right)}$ alone
results to be the complete description of the quantum collapse \cite{key-17}.
But the meaning of $\psi_{i}^{\left(1\right)}$ is that each shell
will condense on the Schwarzschild surface, by stopping the gravitational
collapse \cite{key-17}. Each shell converges to the Schwarzschild
surface and a ``dark star'' forms \cite{key-17}. 

Now, one can find the mass and energy spectra of this ``gravitational
atom'' via a Schrodinger-like approach following \cite{key-19}.
One starts to observe that, if both the shells described by $\psi_{i}^{\left(1\right)}$
converge on the Schwarzschild surface by forming a ``dark star'',
then, by assuming absence of rotations and of dissipation during the
collapse, such a final object will be a spherical symmetric shell.
This is consistent with Einstein's idea in 1939 of the localization
of the collapsing particles within a thin spherical shell \cite{key-18}.
In that case, a ``dark star'' having mass $M$ will be subjected
to the classical potential 
\begin{equation}
V=-\frac{M^{2}}{2R},\label{eq: shell potential}
\end{equation}
which is indeed the self-interaction gravitational potential of a
spherical massive shell, where $R$ is its radius \cite{key-19,key-21}.
In the current case, $R$ is nothing else than the gravitational radius,
which, in a quantum framework, is subjected to quantum fluctuations
\cite{key-22}, due also to the potential absorption of external particles
\cite{key-23}. On the other hand, Eq. ($\ref{eq: shell potential}$)
represents also the potential of a two-particle system composed of
two identical masses $M$ gravitationally interacting with a relative
position $2R$. Hence, the spherical shell becomes physically equivalent
to a two-particle system of two identical masses: but, clearly, as
the shell's mass $M$ does not double, one has to consider the two
identical masses $M$ as being fictitious and representing the real
physical shell. Let us recall the general problem of a two-particle
system where the particles have different masses \cite{key-19,key-20}.
This is a 6-dimensional problem which can be splitted into two 3-dimensional
problems, that of a static or free particle, and that of a particle
in a static potential if the sole interaction which is felt by the
particles is their mutual interaction depending only on their relative
position \cite{key-19,key-20}. One denotes by $m_{1}$, $m_{2}$
the masses of the particles, by $d_{1}$, $d_{2}$ their positions
and by $\overrightarrow{p}_{1}$, $\overrightarrow{p}_{2}$ the respective
momenta. Being $\overrightarrow{d}_{1}-\overrightarrow{d}_{2}$ their
relative position, the Hamiltonian of the system reads \cite{key-19,key-20}
\begin{equation}
H=\frac{p_{1}^{2}}{2m_{1}}+\frac{p_{2}^{2}}{2m_{2}}+V(\overrightarrow{d}_{1}-\overrightarrow{d}_{2}).\label{eq: Hamiltonian 2 particles}
\end{equation}
One sets \cite{key-19,key-20}
\begin{equation}
\begin{array}{ccccc}
m_{T}=m_{1}+m_{2}, &  & \overrightarrow{D}=\frac{m_{1}\overrightarrow{d}_{1}+m_{2}\overrightarrow{d}_{2}}{m_{1}+m_{2}}, &  & \overrightarrow{p}_{T}=\overrightarrow{p}_{1}+\overrightarrow{p}_{2},\\
\\
m=\frac{m_{1}m_{2}}{m_{1}+m_{2}} &  & \overrightarrow{d}=\overrightarrow{d}_{1}-\overrightarrow{d}_{2} &  & \overrightarrow{p}=\frac{m_{1}\overrightarrow{p}_{1}+m_{2}\overrightarrow{p}_{2}}{m_{1}+m_{2}}.
\end{array}\label{eq: sets}
\end{equation}
The change of variables of Eq. ($\ref{eq: sets}$) is a canonical
transformation because it conserves the Poisson brackets \cite{key-19,key-20}.
According to the change of variables of Eq. ($\ref{eq: sets}$), the
motion of the two particles is interpreted as being the motion of
two fictitious particles: i) \emph{center of mass,} having position
$\overrightarrow{D}$, total mass $m_{T}$ and total momentum $\overrightarrow{p}_{T}$
and, ii) the \emph{relative particle} (which is the particle associated
with the relative motion), having position $\overrightarrow{d}$,
mass $m,$ called \emph{reduced mass}, and momentum $\overrightarrow{p}$
\cite{key-19,key-20}. The Hamiltonian of Eq. ($\ref{eq: Hamiltonian 2 particles}$),
considered as a function of the new variables of Eq. ($\ref{eq: sets}$),
becomes \cite{key-19,key-20} 
\begin{equation}
H=\frac{p_{T}^{2}}{2m_{T}}+\frac{p^{2}}{2m}+V(\overrightarrow{d}).\label{eq: Hamiltonian separated}
\end{equation}
The new variables obey the same commutation relations as if they should
represent two particles of positions $\overrightarrow{D}$ and $\overrightarrow{d}$
and momenta $\overrightarrow{p}_{T}$ and $\overrightarrow{p}$ respectively
\cite{key-19,key-20}. The Hamiltonian of Eq. ($\ref{eq: Hamiltonian separated}$)
can be considered as being the sum of two terms \cite{key-19,key-20}:
\begin{equation}
H_{T}=\frac{p_{T}^{2}}{2m_{T}},\label{eq: Hamiltonian 1}
\end{equation}
and 
\begin{equation}
H_{m}=\frac{p^{2}}{2m}+V(\overrightarrow{d}).\label{eq: Hamiltonian 2}
\end{equation}
The term of Eq. ($\ref{eq: Hamiltonian 1}$) depends only on the variables
of the center of mass, while the term of Eq. ($\ref{eq: Hamiltonian 2}$)
depends only on the variables of the relative particle. Thus, the
Schrodinger equation in the representation $\overrightarrow{D},\:\overrightarrow{d}$
is \cite{key-19,key-20}:
\begin{equation}
\left[\left(-\frac{1}{2m_{T}}\triangle_{D}\right)+\left(-\frac{1}{2m}\triangle_{d}+V(d)\right)\right]\Psi\left(D,\:d\right)=E\Psi\left(D,\:d\right),\label{eq: Schrodinger equation two particles}
\end{equation}
being $\triangle_{\overrightarrow{D}}$ and $\triangle_{\overrightarrow{d}}$
the Laplacians relative to the coordinates $\overrightarrow{D}$ and
$\overrightarrow{d}$ respectively. Now, one observes that the reduced
mass of the previously introduced two-particle system composed of
two identical masses $M$ is 
\begin{equation}
m=\frac{M*M}{M+M}=\frac{M}{2}.\label{eq: massa ridotta}
\end{equation}
In that case, by recalling that in Schwarzschild coordinates the BH
center of mass coincides with the origin of the coordinate system
and with the replacements 
\begin{equation}
d\rightarrow2R,\label{eq: replacement}
\end{equation}
the Schrodinger equation ($\ref{eq: Schrodinger equation two particles}$)
becomes 
\begin{equation}
\left(-\frac{1}{2m}\triangle_{2R}+V(2R)\right)\Psi\left(2R\right)=E\Psi\left(2R\right).\label{eq: Schrodinger equation membrana}
\end{equation}
Setting 
\begin{equation}
r\equiv\frac{R}{2},\label{eq: setting}
\end{equation}
the potential of Eq. ($\ref{eq: shell potential}$) becomes 
\begin{equation}
V=-\frac{m^{2}}{r},\label{eq: energia potenziale membrana}
\end{equation}
and the Schrodinger equation in the representation $\overrightarrow{D}=0,\:\overrightarrow{r}$
becomes 
\begin{equation}
\left(-\frac{1}{2m}\triangle_{r}+V(r)\right)\Psi\left(r\right)=E\Psi\left(r\right),\label{eq: rappresentazione r}
\end{equation}
that is 
\begin{equation}
-\frac{1}{2m}\left(\frac{\partial^{2}\Psi}{\partial r^{2}}+\frac{2}{r}\frac{\partial\Psi}{\partial r}\right)+V\Psi=E\Psi.\label{eq: Schrodinger membrana ritrovata}
\end{equation}
 The Schrodinger equation ($\ref{eq: Schrodinger membrana ritrovata}$)
is formally identical to the traditional Schrodinger equation of the
$s$ states ($l=0$) of the hydrogen atom which obeys to the Coulombian
potential \cite{key-19,key-20}
\begin{equation}
V(r)=-\frac{e^{2}}{r}.\label{eq: energia potenziale atomo idrogeno}
\end{equation}
In the potential of Eq. ($\ref{eq: energia potenziale membrana}$)
the squared electron charge $e^{2}$ is replaced by the squared reduced
mass $m^{2}.$ Thus, Eq. ($\ref{eq: Schrodinger membrana ritrovata}$)
can be interpreted as the Schrodinger equation of a particle, the
``electron'', which interacts with a central field, the ``nucleus''.
On the other hand, this is only a mathematical artifact because the
real nature of the quantum BH is in terms of Vaz's shell. For the
bound states ($E<0$) the energy spectrum is 
\begin{equation}
E_{n}=-\frac{m^{5}}{2n^{2}}.\label{eq: spettro energia}
\end{equation}
 Hence, in order to completely solve the problem, one must find the
relationship between the reduced mass and the total energy of Vaz's
shell. This relationship has been found in \cite{key-19} as 
\begin{equation}
E=-\frac{m}{2}.\label{eq: energia totale reale}
\end{equation}
By inserting this last equation in Eq. ($\ref{eq: spettro energia}$),
a bit of algebra permits to obtain the energy spectrum 
\begin{equation}
E_{n}=-\frac{1}{2}\sqrt{n},\label{eq: BH energy levels finale.}
\end{equation}
 and the corresponding mass spectrum 
\begin{equation}
M_{n}=2\sqrt{n}.\label{eq: spettro massa BH finale}
\end{equation}

\section{TIME EVOLUTION OF EINSTEIN-VAZ SCHELLS }

The horizon's absence implies that Einstein-Vaz schells cannot emit
radiation via the Hawking mechanism of pair production from quantum
fluctuations. Hence, Einstein-Vaz schells should emit radiation like
the other bodies. Following \cite{key-24}, from the quantum mechanical
point of view, one physically interprets this radiation as energies
of quantum jumps among the unperturbed levels of Eq. $(\ref{eq: BH energy levels finale.}$).
In quantum mechanics, time evolution of perturbations can be described
by an operator \cite{key-24} 

\emph{
\begin{equation}
U(t)=\begin{array}{c}
W(t)\;\;\;for\;0\leq t\leq\tau\\
0\;\;\;for\;t<0\;and\;t>\tau.
\end{array}\label{eq: perturbazione}
\end{equation}
}Then, the complete (time dependent) Hamiltonian is described by the
operator 

\begin{equation}
H(r,t)\equiv H_{0}(r)+U(t),\label{eq: Hamiltoniana completa}
\end{equation}
where $H_{0}(r)$ is the (time independent) Hamiltonian of the Schrodinger
equation ($\ref{eq: Schrodinger membrana ritrovata}$). Thus, considering
a wave function $\psi(r,t),$ one can write the correspondent \emph{time
dependent Schroedinger equation} for the system 

\begin{equation}
i\frac{d|\psi(r,t)>}{dt}=\left[H_{0}(r)+U(t)\right]|\psi(r,t)>=H(r,t)|\psi(r,t)>.\label{eq: Schroedinger equation}
\end{equation}
 The state which satisfies Eq. ($\ref{eq: Schroedinger equation}$)
is 
\begin{equation}
|\psi(r,t)>=\sum_{n}a_{n}(t)\exp\left(-iE_{n}t\right)|\varphi_{n}(r)>,\label{eq: Schroedinger wave-function}
\end{equation}
where the $\varphi_{n}(r)$ are the eigenfunctions of the time independent
Schroedinger equation ($\ref{eq: Schrodinger membrana ritrovata}$)
and the $E_{n}$ are the correspondent eigenvalues. In the basis $|\varphi_{n}(r)>$,
the matrix elements of $W(t)$ can be written as \cite{key-24} 
\begin{equation}
W_{ij}(t)\equiv A_{ij}\delta(t),\label{eq: a delta}
\end{equation}
where $W_{ij}(t)=<\varphi_{i}(r)|W(t)|\varphi_{j}(r)>$ and the $A_{ij}$
are real. In order to solve the complete quantum mechanical problem
described by the operator ($\ref{eq: Hamiltoniana completa}$), one
needs to know the probability amplitudes $a_{n}(t)$ due to the application
of the perturbation described by the time dependent operator ($\ref{eq: perturbazione}$),
which represents the perturbation associated with the emission of
a particle. For $t<0,$ i.e. before the perturbation operator ($\ref{eq: perturbazione}$)
starts to work, the system is in a stationary state $|\varphi_{n_{1}}(t,r)>,$
at the quantum level $n_{1},$ with energy $E_{n_{1}}=-\frac{1}{2}\sqrt{n_{1}},$
given by Eq. ($\ref{eq: BH energy levels finale.}$). Thus, in Eq.
($\ref{eq: Schroedinger wave-function}$) only the term 
\begin{equation}
|\psi_{n_{1}}(r,t)>=\exp\left(-iE_{n_{1}}t\right)|\varphi_{n_{1}}(r)>,\label{eq: Schroedinger wave-function in.}
\end{equation}
is not null for $t<0.$ This implies $a_{n}(t)=\delta_{nn_{1}}\:\:$
for $\:t<0.$ When the perturbation operator ($\ref{eq: perturbazione}$)
stops to work, i.e. after the emission, for $t>\tau$ the probability
amplitudes $a_{n}(t)$ return to be time independent, having the value
$a_{n_{1}\rightarrow n}(\tau)$. In other words, for $t>\tau\:$ the
system is described by the \emph{wave function} $\psi_{final}(r,t),$
which corresponds to the state 
\begin{equation}
|\psi_{final}(r,t)>=\sum_{n=1}^{n_{1}}a_{n_{1}\rightarrow n}(\tau)\exp\left(-iE_{n}t\right)|\varphi_{n}(x)>.\label{eq: Schroedinger wave-function fin.}
\end{equation}
Therefore, the probability to find the system in an eigenstate having
energy $E_{n}=-\sqrt{n}$, with $n<n_{1}$ for emissions, is given
by 

\begin{equation}
\Gamma_{n_{1}\rightarrow n}(\tau)=|a_{n_{1}\rightarrow n}(\tau)|^{2}.\label{eq: ampiezza e probability}
\end{equation}
By using a standard analysis \cite{key-24}, one obtains the following
differential equation from Eq. ($\ref{eq: Schroedinger wave-function fin.})$ 

\begin{equation}
i\frac{d}{dt}a_{n_{1}\rightarrow n}(t)=\sum_{l=1}^{n}W_{n_{1}l}a_{n_{1}\rightarrow l}(t)\exp\left[i\left(\Delta E_{l\rightarrow n}\right)t\right].\label{eq: systema differenziale}
\end{equation}
 To first order in $U(t)$, by using the Dyson series \cite{key-25},
one gets the solution 
\begin{equation}
a_{n_{1}\rightarrow n}=-i\int_{0}^{t}\left\{ W_{nn_{1}}(t')\exp\left[i\left(\Delta E_{n_{1}\rightarrow n}\right)t'\right]\right\} dt'.\label{eq: solution}
\end{equation}
By inserting Eq. ($\ref{eq: a delta}$) in Eq. ($\ref{eq: solution}$)
one obtains 
\begin{equation}
a_{n_{1}\rightarrow n}=iA_{nn_{1}}\int_{0}^{t}\left\{ \delta(t')\exp\left[i\left(\Delta E_{n_{1}\rightarrow n}\right)t'\right]\right\} dt'=\frac{i}{2}A_{nn_{1}}.\label{eq: solution 2}
\end{equation}
In order to find the quantities $\Gamma_{n_{1}\rightarrow n}(\tau)$
one can use a recent remarkable result of Mathur and Mehta \cite{key-26},
who won the third prize in the 2023 Gravity Research Foundation Essay
Competition for having shown the universality of BH thermodynamics.
In fact, they have shown that any Extremely Compact Object (ECO) like
the Einstein-Vaz shells must have the same thermodynamic properties
of standard BHs. As quantum fields just outside the surface of an
ECO have a large negative Casimir energy similar to the BH Boulware
vacuum, then if the thermal radiation emanating from the ECO does
not fill the near-surface region at the local Unruh temperature, the
consequence is that no solution of gravity equations is possible \cite{key-26}.
In particular, any body, included the Einstein-Vaz shells, whose radius
is sufficiently close to the BH radius, will have the same thermodynamic
properties as the semiclassical BH originally considered by Hawking.
This implies that the temperature of an Einstein-Vaz shell having
mass $M$ must be exactly the Hawking temperature of the corresponding
semiclassical BH, that is $T_{H}\equiv\frac{1}{8\pi M}$ \cite{key-27}.
Therefore, the probability of emission of a single photon is \cite{key-27}
\begin{equation}
\Gamma=\exp(-\frac{\omega}{T_{H}})=\exp\left(-8\pi M\omega\right).\label{eq: hawking probability}
\end{equation}
For a transition between two quantum levels $n_{1}$ and $n,$ with
$n<n_{1},$ one consequently finds
\begin{equation}
\Gamma_{n_{1}\rightarrow n}=\exp\left(-8\pi M\omega_{n_{1}n}\right),\label{eq: tasso transizioni}
\end{equation}
where
\begin{equation}
\omega_{n_{1}n}\equiv M_{n_{1}}-M_{n_{1}}=2\left(\sqrt{n_{1}}-\sqrt{n}\right),\label{eq: frequenza transizione}
\end{equation}
and Eq. ($\ref{eq: spettro massa BH finale}$) has been used in the
last passage of Eq. ($\ref{eq: frequenza transizione}$). On the other
hand, an ambiguity is present in Eq. ($\ref{eq: tasso transizioni}$).
Indeed, because of the quantum transition the mass of the Einstein-Vaz
shell varies from an initial value $M_{n_{1}}$ to a final value $M_{n}<M_{n_{1}}.$
Thus, it is not clear which mass must be inserted in Eq. ($\ref{eq: tasso transizioni}$)
between $M_{n_{1}}$ and $M_{n}.$ The solution of this problem has
been found in \cite{key-23}, where it has been rigorously shown,
via Hawking periodicity argument \cite{key-28}, that the correct
value to be inserted in Eq. ($\ref{eq: tasso transizioni}$) is the
average value between $M_{n_{1}}$ and $M_{n},$ 
\[
\frac{M_{n_{1}}+M_{n}}{2}
\]
which indeed represents the dynamical value of the Einstein-Vaz shell
mass during the quantum transition. Hence, Eq. ($\ref{eq: tasso transizioni}$)
becomes 
\begin{equation}
\Gamma_{n_{1}\rightarrow n}=\exp\left[-4\pi\left(M_{n_{1}}^{2}-M_{n}^{2}\right)\right]=\exp\left[-16\pi\left(n_{1}-n\right)\right].\label{eq: tasso transizioni finale}
\end{equation}
Thus, the probability of emission between two arbitrary quantum levels
of an Einstein-Vaz shell characterized by the two principal quantum
numbers $n_{1}$ and $n$ scales like $\exp\left[-16\pi\left(n_{1}-n\right)\right].$
In particular, for $n=n_{1}-1$ the probability of emission has its
maximum value $\sim\exp(-16\pi)$. This means that the probability
is maximum for two adjacent levels, as one intuitively expects. Combining
Eq. ($\ref{eq: solution 2}$) with Eqs. ($\ref{eq: tasso transizioni finale}$)
and ($\ref{eq: ampiezza e probability}$) one gets 

\begin{equation}
\begin{array}{c}
\exp\left[16\pi\left(n-n_{1}\right)\right]=\frac{1}{4}A_{nn_{1}}^{2}\\
\\
A_{nn_{1}}=2\exp\left[8\pi\left(n-n_{1}\right)\right]\\
\\
a_{n_{1}\rightarrow n}=-i\exp\left[8\pi\left(n-n_{1}\right)\right].
\end{array}\label{eq: uguale}
\end{equation}
Then, one gets $A_{nn_{1}}\sim10^{-11}$ for $n=n_{1}-1$, i.e. when
the probability of emission has its maximum value. This implies that
second order terms in $U(t)$ are $\sim10^{-22}$ and can be neglected.
Clearly, for $n<n_{1}-1$, the approximation is better because the
\$A\_\{n\_\{1\}n\}\$ are even smaller than \$10\textasciicircum\{-11\}\$.
Thus, one can write down the final form of the ket representing the
state as 

\begin{equation}
|\psi_{final}(r,t)>=\sum_{n=1}^{n_{1}}-i\exp\left[8\pi\left(n-n_{1}\right)-iE_{n}t\right]|\varphi_{n}(r)>.\label{eq: Schroedinger wave-function finalissima}
\end{equation}
 The \emph{state} ($\ref{eq: Schroedinger wave-function finalissima}$)
represents a pure final state and the states are written in terms
of an unitary evolution matrix. Consequently, the time evolution of
the Einstein-Vaz shells is unitary as it is requested by a quantum
theory of gravity.

\section{CONCLUSION REMARKS}

In the classical framework we find the lower bounds for the upper
bounds of the deviations of LTB metric from the Riemannian case in
the directions $\frac{\partial}{\partial t}$, $\frac{\partial}{\partial\theta}$,
$\frac{\partial}{\partial r}$, and $\frac{\partial}{\partial\phi}$.
The reader must pay attention to this point that the directions only
determine the linear dynamical systems $L_{\frac{\partial}{\partial t}}$,
$L_{\frac{\partial}{\partial\theta}}$, $L_{\frac{\partial}{\partial r}}$,
and $L_{\frac{\partial}{\partial\phi}}$. In fact these linear dynamical
systems are coordinate free, and the sensitivities and the lower sensitivities
are also coordinate free. Because they depend only to the Levi-Civita
connection determines by LTB metric, which is coordinate free. In
the quantum framework we analysed the properties of the Einstein-Vaz
shells which are the final result of the quantum gravitational collapse
arising from the Lemaitre Tolman Bondi discussed by Vaz in 2014 \cite{key-17}.
In fact, Vaz showed that continued collapse to a singularity can only
be obtained if one combines two independent and entire solutions of
the Wheeler-DeWitt equation. Forbidding such a combination leads naturally
to matter condensing on the Schwarzschild surface during quantum collapse.
In that way, an entirely new framework for BHs has emerged. The approach
of Vaz was also consistent with Einstein's idea in 1939 of the localization
of the collapsing particles within a thin spherical shell \cite{key-18}.
Following \cite{key-19}, we derived the BH mass and energy spectra
via a Schrodinger-like approach, by further supporting Vaz's conclusions
that instead of a spacetime singularity covered by an event horizon,
the final result of the gravitational collapse is an essentially quantum
object, an extremely compact ``dark star''. This ``gravitational
atom'' is held up not by any degeneracy pressure but by quantum gravity
in the same way that ordinary atoms are sustained by quantum mechanics.
Finally, we discussed the time evolution of the Einstein-Vaz shells.

\section*{Competing interests}

This research has been financially supported by Shahid Bahonar University
of Kerman. The Authors declare that there are no potential sources
of conflict of interest in the manuscript.

\end{document}